\documentclass[conference]{IEEEtran}
\usepackage{color}

\ifCLASSINFOpdf
\else
\fi
\usepackage{supertabular}  
\usepackage{afterpage}
\usepackage{multirow}
\usepackage{longtable}
\usepackage{url}
\usepackage[table]{xcolor}
\usepackage{subcaption}
\usepackage{color}
\usepackage{diagbox}
\usepackage{gensymb}
\usepackage{float}
\usepackage{subcaption}
\usepackage{graphicx}
\usepackage{booktabs} 
\usepackage{comment} 
\usepackage{eurosym}

\usepackage{colortbl}
\definecolor{lightblue}{HTML}{CCEEFF}

\usepackage{acronym}
\newacro{AR}[AR]{Augmented Reality}
\newacro{BFT}[BFT]{Byzantine Fault Tolerance}
\newacro{BIIoT}[BIIoT]{Blockchain-based IoT}
\newacro{CPS}[CPS]{Cyber-Physical System}
\newacro{CPPS}[CPPS]{Cyber-Physical Production System}
\newacro{DBFT}[DBFT]{Delegated Byzantine Fault Tolerance}
\newacro{DoF}[DoF]{Degree of Freedom}
\newacro{DoS}[DoS]{Denial of Service}
\newacro{DPoS}[DPoS]{Delegated Proof-of-Stake}
\newacro{ECC}[ECC]{Elliptic Curve Cryptography}
\newacro{ERP}[ERP]{Enterprise Resource Planning}
\newacro{GPU}[GPU]{Graphics Processing Unit}
\newacro{GPUs}[GPUs]{Graphics Processing Units}
\newacro{HMD}[HMD]{Head-Mounted Display}
\newacro{IAR}[IAR]{Industrial Augmented Reality}  
\newacro{ICPS}[ICPS]{Industrial Cyber-Physical System}
\newacro{IMU}[IMU]{Inertial Measuring Unit}
\newacro{IoT}[IoT]{Internet of Things} 
\newacro{IIoT}[IIoT]{Industrial IoT}
\newacro{IVR}[IVR]{Industrial Virtual Reality}
\newacro{M2M}[M2M]{Machine-to-Machine}
\newacro{MES}[MES]{Manufacturing-Execution System}
\newacro{MQTT}[MQTT]{Message Queuing Telemetry Transport}
\newacro{PBFT}[PBFT]{Practical Byzantine Fault Tolerance}
\newacro{P2P}[P2P]{Peer-to-Peer}
\newacro{PLC}[PLC]{Power Line Communication}
\newacro{PLM}[PLM]{Product Lifecycle Management}
\newacro{PoA}[PoA]{Proof-of-Activity}
\newacro{PoB}[PoB]{Proof-of-Burn}
\newacro{PoP}[PoP]{Proof-of-Personhood}
\newacro{PoS}[PoS]{Proof-of-Stake}
\newacro{PoW}[PoW]{Proof-of-Work}
\newacro{NFT}[NFT]{Natural Feature Tracking}
\newacro{RFID}[RFID]{Radio Frequency IDentification}
\newacro{RSA}[RSA]{Rivest–Shamir–Adleman}
\newacro{NSA}[NSA]{National Security Agency}
\newacro{SBC}[SBC]{Single-Board Computer}
\newacro{SCM}[SCM]{Supply Chain Management}
\newacro{SCP}[SCP]{Stellar Consensus Protocol}
\newacro{SDK}[SDK]{Software Development Kit}
\newacro{SDN}[SDN]{Software Defined Networking}
\newacro{SOC}[SoC]{System-on-Chip}
\newacro{TaPoS}[TaPoS]{Transactions as Proof-of-Stake}
\newacro{TLS}[TLS]{Transport Layer Security}
\newacro{VR}[VR]{Virtual Reality}
\newacro{WSN}[WSN]{Wireless Sensor Network}

\begin{document}
%

\title{Design and Validation of a Bluetooth 5\\ Fog Computing-Based Industrial CPS Architecture \\for Intelligent Industry 4.0 Shipyard Workshops}




\author{\IEEEauthorblockN{Paula Fraga-Lamas\IEEEauthorrefmark{1}, Member, IEEE,
Peio Lopez-Iturri\IEEEauthorrefmark{2}\IEEEauthorrefmark{3}, Member, IEEE,\\
Mikel Celaya-Echarri\IEEEauthorrefmark{4}, Student Member, IEEE,
Oscar Blanco-Novoa\IEEEauthorrefmark{1}, 
Leyre Azpilicueta\IEEEauthorrefmark{4}, Member, IEEE,\\
Jos\'e Varela-Barbeito\IEEEauthorrefmark{5},   
Francisco Falcone\IEEEauthorrefmark{2}\IEEEauthorrefmark{3}, Senior Member, IEEE and\\
Tiago~M. Fern\'andez-Caram\'es\IEEEauthorrefmark{1}, Senior Member, IEEE}

\IEEEauthorblockA{\IEEEauthorrefmark{1}Dept. of Computer Engineering, CITIC, Universidade da Coru\~na, 15071, A Coru\~na, Spain;} 
\IEEEauthorblockA{\IEEEauthorrefmark{2}Dept. of Electric, Electronic and Communication Engineering, Public University of Navarre, \mbox{31006, Pamplona, Spain;}}
\IEEEauthorblockA{\IEEEauthorrefmark{3}Institute of Smart Cities, Public University of Navarre, 31006, Pamplona, Navarra, Spain.;}
\IEEEauthorblockA{\IEEEauthorrefmark{4}School of Engineering and Sciences, Tecnologico de Monterrey, 64849, Monterrey, Mexico;}
\IEEEauthorblockA{\IEEEauthorrefmark{5}Navantia S. A., Astillero R\'ia de Ferrol, Taxonera, s/n, Ferrol 15403, Spain. }}

\maketitle

\begin{abstract}
Navantia, one of Europe's largest shipbuilders, is creating a fog computing-based Industrial Cyber-Physical System (ICPS) for remote monitoring in real-time of their pipe workshops. Such a monitoring process, which involves pipe traceability and tracking, is a unique industrial challenge, 
given their metallic content, massive quantity and heterogeneous typology, as well as to the number of complex processes involved.
Pipe improved location, from production and through their lifetime, can provide significant productivity and safety benefits to shipyards and foster innovative applications in process planning.
Bluetooth 5 represents a cost-effective opportunity to cope with this harsh environment, since it has been significantly enhanced in terms of low power consumption, range, speed and broadcasting capacity.
Thus, this article proposes a Bluetooth 5 fog computing-based ICPS architecture that is designed to support physically-distributed and low-latency Industry 4.0 applications that off-load network traffic and computational resource consumption from the cloud.  
In order to validate the proposed ICPS design, one of Navantia's pipe workshops has been modeled through an in-house-developed 3D ray launching radio planning simulator that considers three main intrinsic characteristics: the number of pipes, the main working areas with their corresponding machines, and the daily workforce. The radio propagation results obtained by the simulation tool are validated through empirical measurements. These results aim to provide guidelines for ICPS developers, network operators and planners to investigate further complex industrial deployments based on Bluetooth 5.
\end{abstract}


%
\IEEEpeerreviewmaketitle

\section{Introduction}

The application of the Industry 4.0 and Industrial Internet of Things (IIoT) paradigms to traditional industrial facilities is changing dramatically the way factories and industries operate and communicate thanks to the use of some of the latest technologies for managing, monitoring and optimizing industrial processes.
Shipbuilding is one of the many industries that can benefit from Industry 4.0 and IIoT principles, since building large vessels is a really complex task that involves many processes that can be enhanced and optimized through technology to meet time and quality constraints.

Navantia, a Spanish 300-year old company, is one of the shipbuilders that is pushing the application of Industry 4.0 technologies to improve its competitiveness. To do so, Navantia is leading the ``Shipyard 4.0'' project, which seeks to create a modern shipyard through the application of the Industry 4.0 principles to build the next generation of hi-tech military and civil vessels.

 Among the different research lines established within the Shipyard 4.0 project, the so-called Auto-ID line is arguably the one with a major impact on a shipyard, since its objective, which consists in identifying objects, products, facilities, tools or people automatically throughout their lifetime, has potential to impact  almost every shipbuilding process. For the sake of brevity and clarity, this article will be focused on the pipe workshop, which is one of the most important locations in every Navantia's shipyard because of the number of pipes to be built (between 15,000 and 40,000), their different features (e.g., dimensions, material, accessories) and the fact that each pipe requires to be processed through different phases depending on its specific features.
 
\begin{figure*}[!hbt]
        \centering
		\includegraphics[width=1\linewidth]{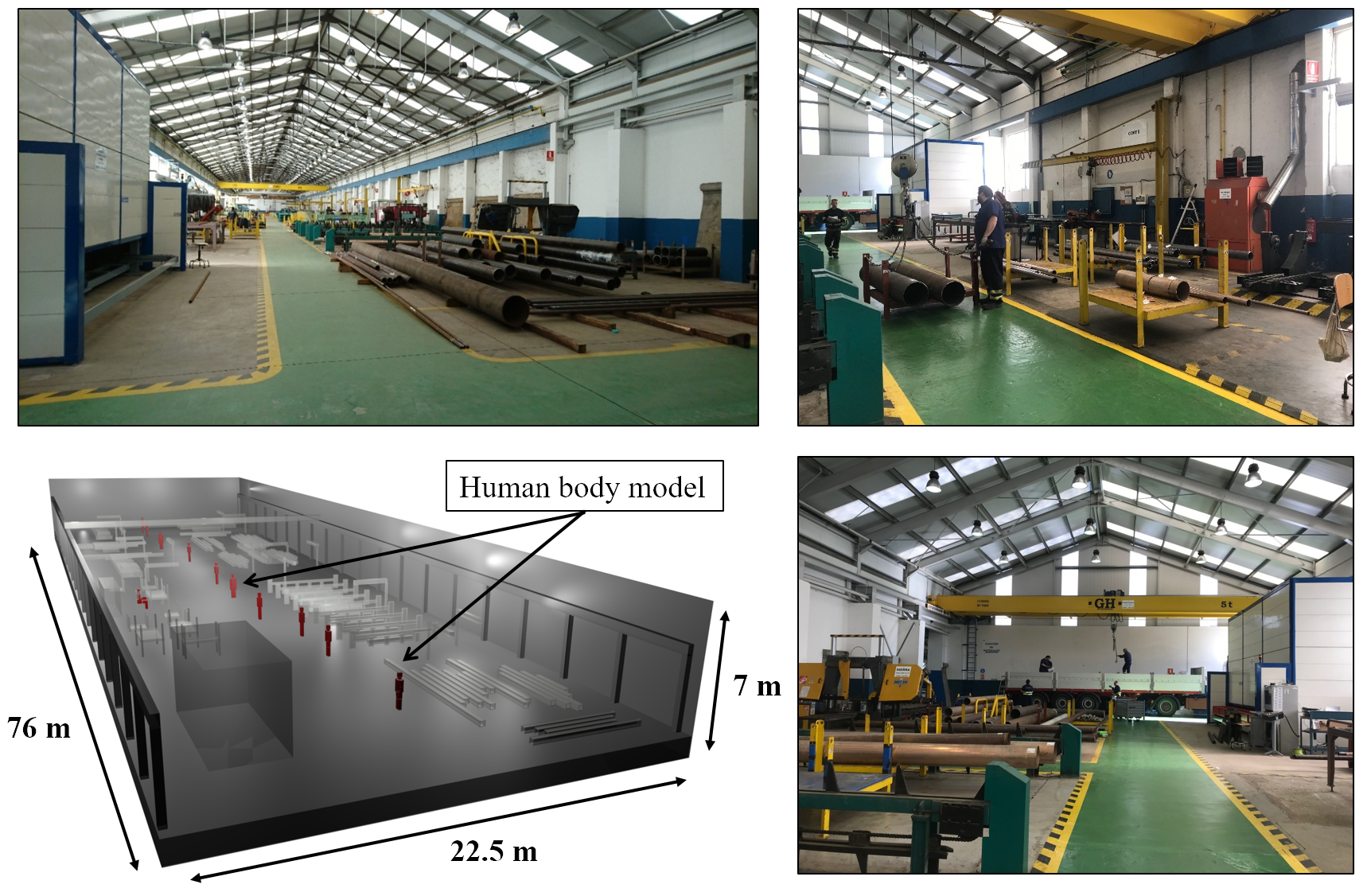}
		\caption{Pictures of the general view (top left) and different areas (top right, bottom right) of the pipe workshop together with simulation scenario (bottom left).} 
		\label{figure:montajePeio}
\end{figure*}

The specific pipe workshop where the experiments shown in this paper will take place is an unobstructed two-wing building that is roughly 120\,m long and 40\,m wide. Part of the workshop is shown in Figure \ref{figure:montajePeio} together with its 3D model. As it can be seen, the shipyard contains large metallic objects that impact wireless communications significantly \cite{RSS}. Other factors that characterize the scenario are a high relative humidity level, since the workshop is located by the sea, and high temperatures in certain areas.
Inside the building, pipes go through multiple areas depending on the manufacturing processes they are subjected to (such processes are briefly  described later in Section \ref{sec:lifecycle}).


In our previous work, an \ac{ICPS} was proposed to track pipes in real time \cite{SmartPipes}. Such an ICPS made use of active \ac{RFID} for locating and identifying pipes in highly-metallic shipyard environments. Nonetheless, the relatively high cost of the deployment (each of the selected RFID tags costs around \euro\,30, while RFID readers around \euro\,2,000), prevents their use in certain applications and industries. In addition, due to the progressive growth in the number of products to be monitored (whose communication transactions impact the remote cloud performance) and the need for location awareness and low-latency responses, a fog-computing architecture was designed and tested in the shipyard \cite{FogAutoID}.

In such a scenario, wireless communications are constrained to specific conditions, such as strong multipath components, owing to highly reflective surroundings or high levels of electromagnetic interference \cite{Stenumgaard2013}, which can impact system reliability \cite{NIST1951}. In the past, intensive measurement campaigns were performed, showing the relevance of multiple interference sources (e.g., power converters, welding systems, electrical engines) and the influence of the objects present in the industrial environment \cite{NIST1951}, providing radio planning solutions based on regressive propagation loss models \cite{NIST1951} or the use of stochastic models for short range wireless communications \cite{Savazzi2014}. The proposed methods provide an estimation on the initial coverage/capacity as a function of the required sensitivity, but provide little information in terms of time domain characterization, such as power delay profiles and delay dispersion.

Moreover, the communication technologies employed so far to cope with IIoT scenarios have not yet been able to manage properly the requirements of long reading range and cost-efficiency, specially in scenarios that are hostile for electro-magnetic propagation. Nevertheless, recent solutions like Low-Power Wide Area Networks (LPWANs) have emerged as a promising alternative that can be combined with the so-called short-range technologies. This is one of the reasons why, in 2016, the Bluetooth Special Interest Group (SIG) presented Bluetooth 5 \cite{Collotta2018}, whose primary aim is to provide significant enhancements with respect to the preceding specifications regarding range, speed, broadcasting capacity, reduced power consumption and coexistence with other cellular and LPWAN technologies. 

Specifically, the pipe identification system described in this article makes use Bluetooth 5.0 devices, since the latest core specification (Bluetooth 5.1) is still very recent (it was released on January 21st 2019 \cite{SIG}), so, as of writing, there is no Bluetooth 5.1 hardware available. Bluetooth 5.0 has been devised for being used in \ac{IoT} applications, it is able to provide long reading ranges in harsh environments (in terms of communications propagation) and its cost is expected to decrease progressively (it is still in its earliest commercialization stages) as Bluetooth 5.0 devices become massively produced. In addition, this article analyzes the feasibility of the use of Bluetooth 5.0 in a pipe workshop, validating it through simulated models and practical experiments. This validation has been performed by means of an in-house developed 3-Dimensional Ray Launching (3D RL) simulation algorithm. To the knowledge of the authors,
this deterministic approach has not been used previously in similar industrial scenarios. Moreover, it is also the first time that the algorithm has been validated for Bluetooth 5.0 communications. Furthermore, no previous evaluation has been found in the literature on the use of Bluetooth 5.0 for similar scenarios.

The remainder of this paper is organized as follows. 
Section \ref{sec:relatedWork} describes the life-cycle of a pipe in Navantia's pipe workshop and reviews the most relevant ICPSs and Bluetooth 5.0 developments for industrial scenarios.
Section \ref{sec:design} details the proposed system architecture and the characteristics of the Bluetooth 5.0 devices used.
Section \ref{sec:experiments} analyzes the radio propagation characteristics of the proposed Bluetooth 5.0 system in Navantia's pipe workshop by comparing the results obtained by the in-house simulation tool and empirical measurements.
Finally, Section \ref{sec:Conclusions} is devoted to the conclusions.

\section{Related Work} \label{sec:relatedWork}

\subsection{Basic operation of the pipe workshop} \label{sec:lifecycle}
Although the different processes along a pipe life-cycle take place in several scenarios throughout the shipyard (e.g., in the block outfitting workshop or inside the ship), the most important ones related to their manufacturing occur in the pipe workshop. Every pipe workshop is usually divided in several functional areas where different processes occur, like cutting, bending, accessory fitting, welding or cleaning. 

As raw pipes arrive to the workshop, they are stored in a reception area according to their typology and the production needs. In the first processing stage, pipes are cut with high-precision saws following the engineering requirements and an identification tag is attached. Some of the pipes are bent through computer-aided bending machines to adapt them to their final shape. Before further processing the pipes, they are cleaned and degreased. In the so-called manufacturing areas, operators add accessories (e.g., hydraulic valves, connection fittings) or weld the different sub-pipes that conform the final pipe. After welding, pipes are again cleaned and rinsed in bathtubs with hot water, acids, caustic solutions and pressurized water. Finally, pipes are stacked on pallets and moved by cranes through the workshop to different storage areas. 


\subsection{ICPS and fog computing for shipyards}
ICPSs are expected to empower the fourth industrial revolution by enabling the creation, operation and interconnection of intelligent heterogeneous systems \cite{Jiang2018}. Although cloud and service-based computing are the traditional paradigms implemented by ICPSs \cite{Cheng2018}, they get in conflict with the Industry 4.0 requirements related to the dependency on external connectivity, the need for geographically distributed heterogeneous platforms, decentralized processing, autonomous decision-making, single point of failure, scalability and reliable real-time control of critical resources. 

In order to confront these issues, fog computing solutions have been introduced as they provide storage and local processing together with low latencies and enhanced security by pushing the computation resources onto the network edge \cite{Donovan2018}. Although there are a number of ICPS for different industries, there are few examples of ICPS for shipyards in the literature \cite{FogAutoID} and apart from the previous work of the authors, none of them deals with the use of fog computing to fulfill the requirements of shipyards.
In addition, there are not many recent references in the literature on the use of Bluetooth in shipyards. For instance, in 2006 the authors of \cite{Kawakubo2006} proposed a Bluetooth system for positioning workers. The authors created a mockup workshop and tested the system, achieving a 1.2\,m accuracy in a cluttered environment.

\begin{figure*}[!htb]
        \centering
		\includegraphics[width=0.77\linewidth]{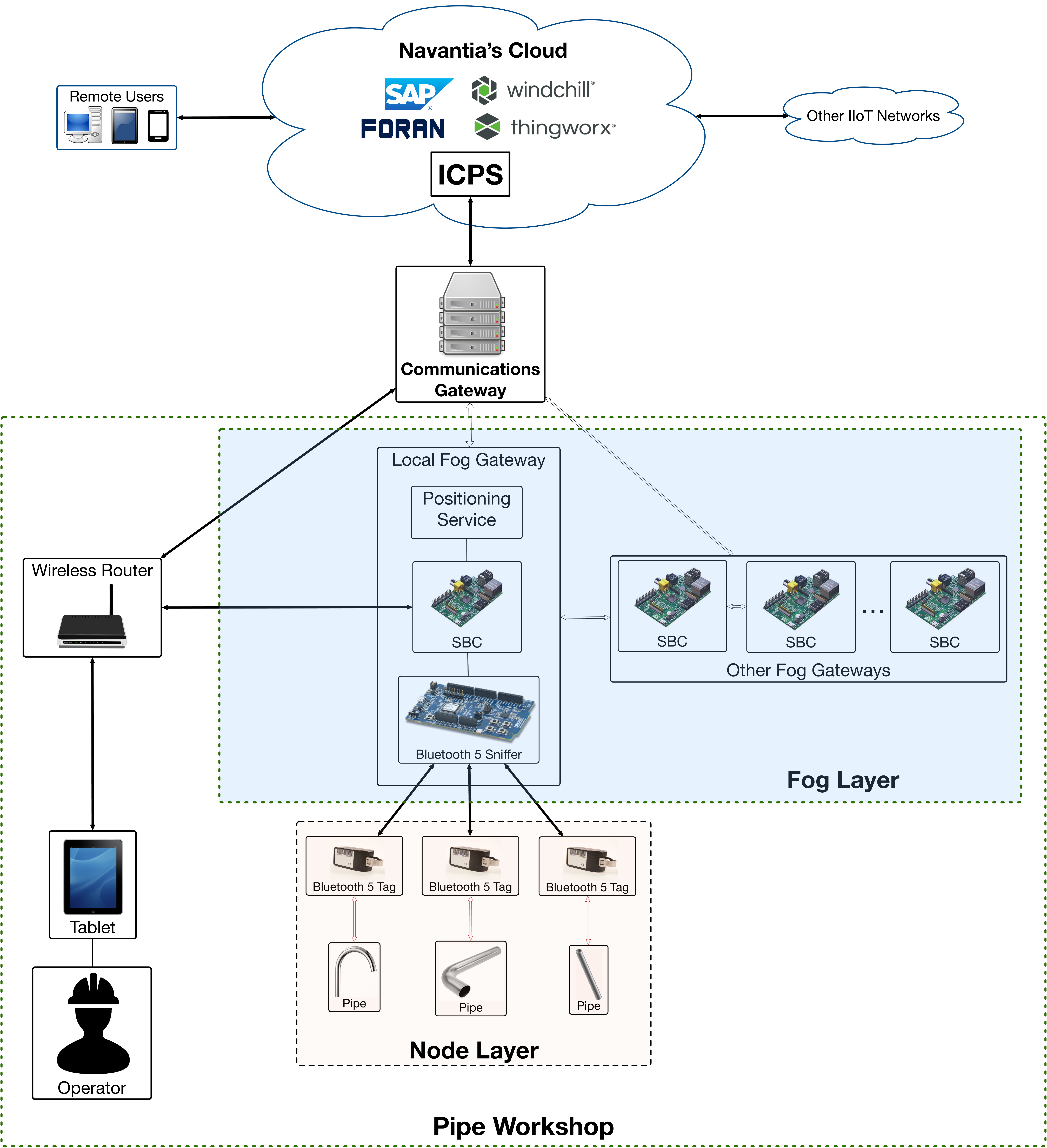}
		\caption{Proposed communications architecture.} 
		\label{figure:architecture}
\end{figure*}

\section{Design of the system} \label{sec:design}

\subsection{Communications architecture} 

Fig. \ref{figure:architecture} shows the proposed communications architecture, which is divided into three layers. The bottom layer is composed by pipes that are attached to Bluetooth 5 tags. The signal strength of the tags is read by fog gateways that are connected to Bluetooth 5 sniffers. A fog gateway is essentially a \ac{SBC} (e.g., a Raspberry Pi) that runs the positioning service and that provides fast responses to the operators. Fog gateways are scattered throughout a shipyard and are able to communicate with each other in order to collaborate when providing services. Fog gateways also communicate with the top layer, which is a cloud where the most compute-intensive tasks are executed by either proprietary developments (e.g., the developed ICPS) or third-party software (SAP is used as \ac{ERP}, FORAN is used for ship design, Windchill works as \ac{PLM} and ThingWorx is being tested as IIoT platform).

\begin{figure*}[!hbt]
        \centering
		\includegraphics[scale=0.33]{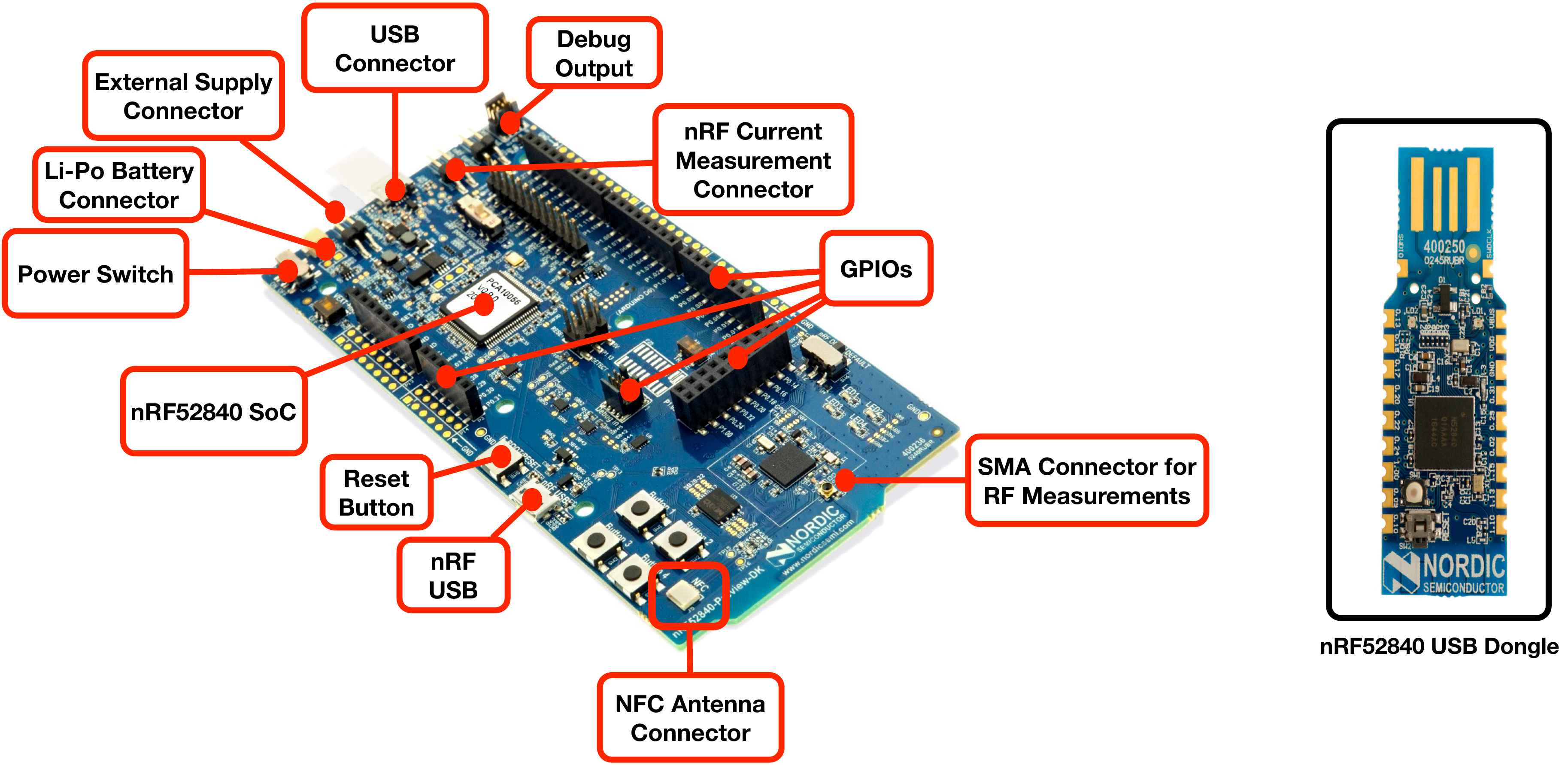}
		\caption{Internal components of the Bluetooth 5 tag (left) and sniffer (right).} 
		\label{figure:PDK}
\end{figure*}

\subsection{Bluetooth 5.0 Sniffer and Tags}  \label{sec:Applications}    

There are different wireless technologies that can be used for providing communications interfaces and identification to the pipe tags (the most relevant are compared in Table \ref{table:protocols}). Among them, there are several that seem promising, like Wi-Fi HaLow or NB-IoT, but, as of writing, they are only supported by a few devices. Although Bluetooth 5 is still being widely adopted by hardware manufacturers, its low power consumption, long range and backward compatibility with other Bluetooth specifications make it a good choice for pipe tracking in harsh industrial environments. Specifically, in the experiments presented in this article, each tag implements Bluetooth 5.0 through an nRF52840 Preview Development Kit, which currently costs around \euro\,35 and is based on an nRF52840 \ac{SOC}. The kit is not only able to be programmed to act as a Bluetooth 5.0 device, but also as an ANT, ANT+ or NFC device. As it can be observed on the left hand side in Figure \ref{figure:PDK}, the kit contains different connectors for communications, debugging and powering the board, as well as multiple General-Purpose Input-Output (GPIO) pins. 

Regarding the Bluetooth 5.0 reader (actually, Bluetooth 5.0 sniffer in Figure \ref{figure:architecture}), an nRF52840 dongle was selected (it is shown in Figure \ref{figure:PDK} on the right). Such a dongle is a small and low-cost (less than \euro\,5 as of writing) USB device whose firmware can be reprogrammed to support Bluetooth 5.0, Bluetooth mesh, Thread, ZigBee, 802.15.4, ANT and other 2.4\,GHz proprietary protocols.

\begin{table*}[!hbt]
\caption{Main characteristics of the most relevant communication and identification technologies for pipe tags.}
\def\arraystretch{2} 
\centering
\resizebox{\textwidth}{!}{
        \begin{tabular}{m{2.5cm}<{\centering}m{2cm}<{\centering}m{2.1cm}<{\centering}m{2cm}<{\centering}m{4cm}<{\centering}m{3cm}<{\centering}}
        \toprule
		\textbf{Technology}&\textbf{Operating Frequency}&\textbf{Maximum Range}&\textbf{Speed}&\textbf{Main Characteristics}&\textbf{Main Applications}\\  
		\midrule
\rowcolor{lightblue}  Bluetooth 5 &2.4\,GHz &$<$250\,m& up to 2\,Mbit$/$s& Low power (batteries can last days to weeks)&Beacons, IoT applications\\
DASH7/ISO 18000-7 &315--915\,MHz&$<$10\,km& 27.8\,kbit$/$s& Low power (batteries can last months to years)&Product tracking and identification\\
\rowcolor{lightblue} HF RFID &3--30\,MHz (13.56\,MHz) & $<$10\,m&$<$640\,kbit$/$s &Low cost readers and tags & Product tracking, payments\\
IQRF& 868\,MHz&hundreds of meters& 100\,kbit/s & Low power and long range&IoT applications\\ 
\rowcolor{lightblue} LF RFID &30--300\,KHz (125\,KHz) &$<$10\,cm&$<$640\,kbit$/$s& Low cost readers and tags& Product tracking and security access controls\\
NB-IoT &LTE in-band, guard-band&$<$35\,km&$<$250\,kbit$/$s&Low power and wide area coverage&IoT applications\\
\rowcolor{lightblue} NFC &13.56\,MHz &$<$20\,cm&  424\,kbit/s & Low cost, tags require no batteries & Asset tracking, payments\\
RuBee&131\,KHz&30\,m&  8\,kbit/s  & Magnetic propagation&Applications with harsh electromagnetic propagation\\    
\rowcolor{lightblue} LoRa, LoRaWAN&2.4\,GHz&kilometers&0.25$-$50\,kbit/s  & Low power and wide range&IoT applications\\
SigFox&868-902\,MHz&50\,km& 100\,kbit/s & Global cellular network&IoT applications\\ 
\rowcolor{lightblue} UHF RFID &30\,MHz--3\,GHz   &  $<$120\,m& $<$640\,kbit$/$s& Low cost& Asset tracking\\
UWB/IEEE 802.15.3a&3.1 to 10.6\,GHz&$<$\,10\,m&$>$110\,Mbit$/$s & Low power (batteries last from hours to days)& Real-Time Location Systems, short-distance streaming\\
\rowcolor{lightblue} Weightless-P   &License-exempt sub-GHz&15\,Km& 100\,kbit$/$s& Low power&IoT applications\\
Wi-Fi (IEEE 802.11b/g/n/ac)&2.4--5\,GHz &$<$150\,m&up to 433\,Mbit$/$s (one stream)  & High power consumption (batteries usually last hours)&High-speed, ubiquity\\
\rowcolor{lightblue} Wi-Fi HaLow/IEEE 802.11ah	&868-915\,MHz&$<$1 km& 100\,Kbit/s per channel & Low power &IoT applications\\ 
WirelessHART&2.4\,GHz&$<$10\,m&250\,kbit$/$s &Compatibility with HART protocol&Wireless sensor network applications\\
\rowcolor{lightblue} ZigBee &868-915\,MHz, 2.4\,GHz & $<$100\,m&20$-$250\,kbit$/$s & Low power (batteries last months to years), up to 65,536 nodes & Smart Home and industrial applications\\
       \bottomrule
	\end{tabular}
    }
     \label{table:protocols}
	\end{table*}

\begin{figure*}[!hbt]
        \centering
		\includegraphics[width=1\linewidth]{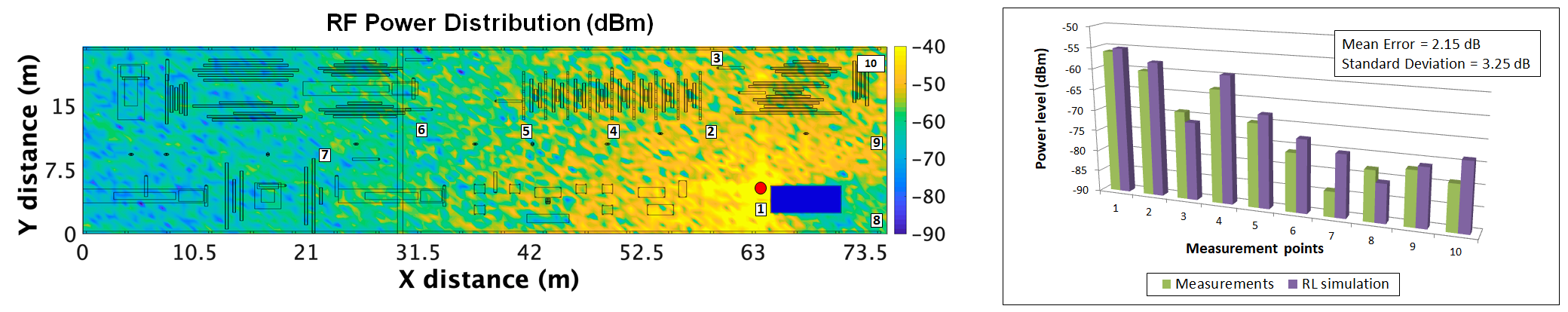} 
		\caption{Bi-dimensional RF power distribution results obtained by the 3D RL at a height of 1.5\,m (left). Transmitter location is represented by a red dot. Comparison between measurements and simulation results (right).}
		\label{figure:figura1aUPNA}
\end{figure*}

\begin{figure*}[!hbt]
        \centering
		\includegraphics[width=1\linewidth]{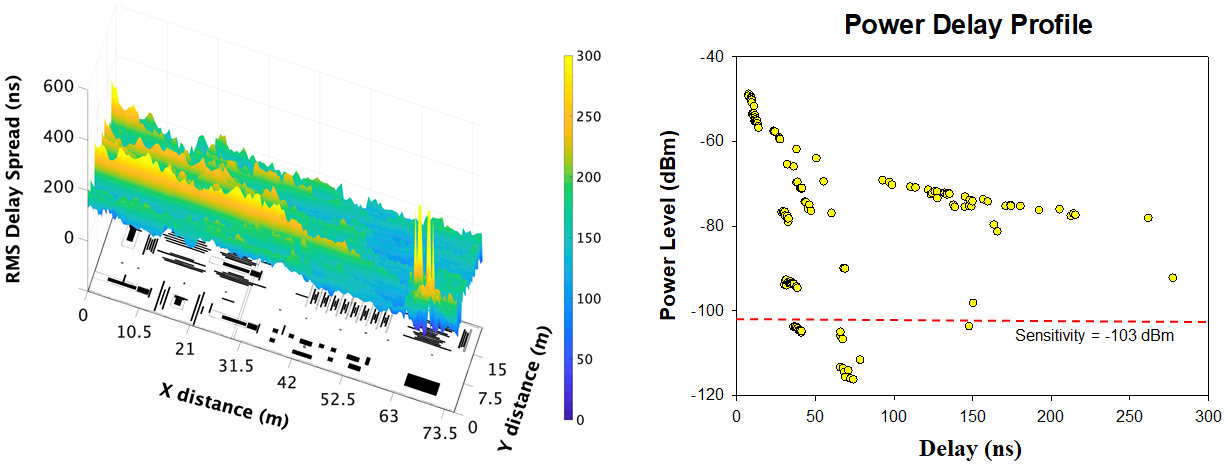} 
		\caption{Delay Spread results for a plane at 1.5\,m height obtained by 3D RL simulations (left). Power Delay Profile corresponding to a point nearby the measurement point 1 (right).}
		\label{figure:figura1bUPNA}
\end{figure*}

\section{Radio propagation assessment of the pipe workshop} \label{sec:experiments}

Due to the particular characteristics and its complexity in terms of morphology, radio propagation within the pipe workshop has been analyzed by a deterministic Ray Launching (RL) simulation tool. Specifically, an in-house developed 3-D Ray Launching algorithm based on Geometrical Optics (GO) and the Uniform Theory of Diffraction (UTD) has been used \cite{Azpilicueta2015}. The GO/UTD technique combination has been validated in the literature to predict wireless propagation within complex 3-D environments \cite{Salous2013,Loredo2008}. Basically, the RL technique is a precise approximation of the full wave methods, which are based on Maxwell's equations. The RL methodology is based on the principle that the wave front of the radiated electromagnetic wave is approximated with a set of rays (launched by a transmitter antenna) that propagate along the full volume of the scenario, following a combination of optics and electromagnetic assumptions. Rays are launched at an elevation angle $\theta$ and an azimuth angle $\emptyset$, considering the spherical coordinate system. When a ray finds an obstacle in its path, the direct, reflected and refracted rays are assessed with GO. The new angles of the reflected and transmitted rays are provided by the Snell’s law application. However, some shadowing problems are presented by applying the GO approach due to field prediction impairments in edges and discontinuities areas. This problem is solved in the algorithm by introducing the diffraction phenomenon based on UTD. In this way, the diffraction coefficients on the edges of the diffractive elements are considered to predict shadowing areas’ field. In order to reduce computational cost and enable analysis capability of large scenarios, hybrid techniques based on Neural Network interpolators, application of Electromagnetic Diffusion Equation and Collaborative Filtering data mining have been included in the in-house simulation code.

The procedure followed for the application of the simulation tool first consists in creating a complete 3-D scenario that considers all the obstacles of the environment under evaluation. As the considered industrial scenario is a complex environment in terms of radio wave propagation (mainly due to the rich multipath components created by the large quantity of metallic obstacles within it), a precise simulation environment is needed in order to obtain accurate estimations. For such a purpose, the different areas of the pipe workshop have been recreated in the simulation tool, taking into account the cranes, the metal beams, the metal cabinets, the stacking areas (which contain metal pipes of different lengths), the pallet area for loading and unloading material, the corridors and a random distribution of operators, both in the corridors and in the different working areas. In the simulation tool, the material properties for all the elements within the scenario are evaluated by means of the dielectric constant and permittivity at the operation frequency range of the wireless system under analysis. In addition, variables such as frequency of operation, maximum number of permitted reflections (i.e., interactions between the propagated ray and obstacles), angular and spatial resolution (i.e., the angle between launched rays and simulations mesh size, respectively), as well as transceiver setups (antenna type, transmitted power level, radiation pattern) are considered as input parameters for simulation. 

During the simulation process, the 3-D scenario is divided into a matrix structure of fixed-size cuboids (defined by the spatial resolution parameter) in such a way that when a ray goes through a specific cuboid, its propagation parameters are stored there. After ray propagation simulation, all the data collected per each cuboid can be retrieved for subsequent analysis. The estimated main result is the received power at each cuboid, which is calculated as the sum of incident electric vector fields (magnitude and phase) in an interval of time $\Delta$t defined by the user and dependent of the transmission data rate of the wireless communication.

\subsection{3-D Ray Launching results}
The presented in-house 3-D RL tool has been used for the radio propagation assessment within the pipe workshop. For the simulations, the aforementioned variables have been defined based on previous convergence analysis studies and the real characteristics of the employed Bluetooth 5.0 devices:
\begin{itemize}
    \item Operating frequency: 2.44\,GHz.
    \item Antenna type: Monopole.
    \item Transmitted power level: 0 dBm.
    \item Permitted maximum reflections: 6.
    \item Horizontal/vertical angular resolution of the launched rays: $\pi$/180$\,rad$.
    \item Cuboid size: 0.5\,m\,$\times$\,0.5\,m\,$\times$\,0.5\,m.
\end{itemize}

RF power distributions as well as time-domain results for the whole volume of the scenario have been obtained. Since the presence of human beings within the scenario is common, a computational human body model has been included for the simulations (previously illustrated in Figure \ref{figure:montajePeio}).

Figure \ref{figure:figura1aUPNA}, on the left, shows the bi-dimensional RF power distribution at a height of 1.5\,m. In such a Figure, the location of the Bluetooth 5.0 tag, which acts as a beacon (at a height of 1.62\,m), is represented by a red dot. The measurement points (with heights from 0.1\,m to 0.9\,m) are depicted as white rectangles that are numbered from 1 to 10. 

Figure \ref{figure:figura1aUPNA}, on the right, shows a comparison between the obtained simulation results and the empirical measurements. Such results show the expected fast signal variations due to the constructive and destructive effects of the multipath propagation components. The obtained mean error between measurements and simulation estimations is 2.15\,dB with a standard deviation of 3.25\,dB, which is a low error taking into account that the performed simulations are static, but the real scenario is dynamic, with operators, machinery and even vehicles moving within it. Therefore, the obtained simulation results have been considered a good approximation and in the same way, the simulation tool validated for its use in this specific environment. 

In addition, Figure \ref{figure:figura1bUPNA} shows the obtained time-domain radio propagation results, which show the great complexity this industrial environment has in terms of multipath propagation. On the left of the Figure, Delay Spread results for the plane at 1.5\,m height can be seen. These results, presented in nanoseconds, express the time interval between the first and the last received multipath component at each point of the scenario (in this case, the points of the bi-dimensional plane at 1.5\,m height). Besides from the typical values nearby the Bluetooth 5.0 tag, large values can also be observed at points further away from it, showing that the scenario under analysis is very rich in terms of multipath propagation (i.e., with a large amount of multipath components that propagate within it). This phenomenon can be clearly seen in the graph depicted on the right side of Figure \ref{figure:figura1bUPNA}, where the Power Delay Profile (PDP) for a point nearby the measurement point 1 is shown. The PDP shows all the multipath components (represented by yellow dots) that reached a specific point within the simulated scenario. 

All the obtained simulation results show the great complexity of this specific industrial scenario, which are in accordance with the results and conclusions of the NIST Technical Note 1951 \cite{NIST1951}. This Technical note also emphasizes that accurate tools are needed in order to perform optimized radio planning tasks for industrial environments. In this way, the presented deterministic 3-D RL simulation tool can aid satisfactorily in these radio planning tasks.

\section{Conclusions} \label{sec:Conclusions}
Pipe traceability is a unique industrial challenge in shipbuilding. To tackle this issue, this article presented the design and validation of a Bluetooth 5.0 fog computing-based ICPS architecture for a real pipe workshop. The proposed architecture enables novel Industry 4.0 applications and guarantees quick responses to critical events while forwarding complex and computing-intensive tasks to the cloud. To implement the architecture the latest and enhanced version of Bluetooth, Bluetooth 5.0, was considered as the best cost-effective technology in order to develop the proposed ICPS. 
Taking into account the harsh electromagnetic conditions in the pipe workshop, a radio propagation assessment was performed using an in-house 3-D Ray Launching simulation tool. A 3-D pipe workshop scenario was modeled considering the number of pipes and the main working areas, including specific machinery and the daily workforce. 
The results obtained with the simulation tool showed good overall accuracy when compared with empirical measurements, therefore validating the use of the tool for such an environment.
As a consequence, it can be stated that the simulation tool is able to provide useful guidelines during the network planning phases for the future deployment of Bluetooth 5 based systems in intelligent Industry 4.0 shipyard workshops.

\section*{Acknowledgment}
 This work was supported by the Pipe Auto-ID research line of the Navantia-UDC Joint Research Unit.
 


\begin{thebibliography}{999}


\bibitem{RSS} P. Fraga-Lamas, T. M. Fern\'andez-Caram\'es, D. Noceda-Davila and M. Vilar-Montesinos, ``RSS stabilization techniques for a real-time passive UHF RFID pipe monitoring system for smart shipyards'', 2017 IEEE International Conference on RFID (RFID), Phoenix, AZ, 2017, pp. 161-166.

\bibitem{SmartPipes}
P. Fraga-Lamas, D. Noceda-Davila, T. M. Fern\'andez-Caram\'es,
M. D\'iaz-Bouza, and M. Vilar-Montesinos, ``Smart pipe system for a shipyard 4.0'', Sensors, vol. 16, no. 12, p. 2186, Dec. 2016

\bibitem{FogAutoID}
T. M. Fern\'andez-Caram\'es, P. Fraga-Lamas, M. Su\'arez-Albela and M. D\'iaz-Bouza 
``A Fog Computing Based Cyber-Physical System for the Automation of Pipe-Related Tasks in the Industry 4.0 Shipyard'', Sensors, vol. 18, no. 6, p. 1961, June 2018.

\bibitem{Stenumgaard2013} P. Stenumgaard, J. Chilo, J. Ferrer-Coll, and P. {\"A}ngskog, ``Challenges and Conditions for Wireless Machine-to-Machine
Communications in Industrial Environments'', IEEE Communications Magazine,
pp. 187-192, June 2013.

\bibitem{NIST1951}     R. Candell et al., ``Industrial Wireless Systems Radio Propagation Measurements'', NIST Technical Note 1951, January 2017,
Available on: \url{https://doi.org/10.6028/NIST.TN.1951} (accessed on 28 February 2019).

\bibitem{Savazzi2014}      S. Savazzi, V.Rampa, and U. Spagnolini, ``Wireless Cloud Networks for the Factory of Things: Connectivity Modeling
and Layout Design'', IEEE Internet Of Things Journal, Vol. 1, No. 2, pp. 180-194, April 2014.

\bibitem{Collotta2018} M. Collotta, G. Pau, T. Talty and O. K. Tonguz, ``Bluetooth 5: A Concrete Step Forward toward the IoT'', in IEEE Communications Magazine, vol. 56, no. 7, pp. 125-131, July 2018.

\bibitem{SIG} Bluetooth SIG Core Specifications. Available online: \url{https://www.bluetooth.com/specifications/bluetooth-core-specification}  ( accessed on 28 February 2019).



\bibitem{Jiang2018}
Y. Jiang, S. Yin and O. Kaynak, ``Data-Driven Monitoring and Safety Control of Industrial Cyber-Physical Systems: Basics and Beyond'', in IEEE Access, vol. 6, pp. 47374-47384, 2018.

\bibitem{Cheng2018} B. Cheng, J. Zhang, G. P. Hancke, S. Karnouskos and A. W. Colombo, ``Industrial Cyberphysical Systems: Realizing Cloud-Based Big Data Infrastructures'', in IEEE Industrial Electronics Magazine, vol. 12, no. 1, pp. 25-35, March 2018.



\bibitem{Donovan2018} P. O'Donovan, C. Gallagher, K. Bruton, D. T.J. O'Sullivan, ``A fog computing industrial cyber-physical system for embedded low-latency machine learning Industry 4.0 applications'', in Manufacturing Letters, vol. 15, Part B, 2018, pp. 139-142.





\bibitem{Kawakubo2006}
S. Kawakubo et al., ``Wireless network system for indoor human positioning'', 2006 1st International Symposium on Wireless Pervasive Computing, Phuket, 2006, pp. 6. 


\bibitem{Azpilicueta2015} L. Azpilicueta, P. L\'opez-Iturri, E. Aguirre, J. J. Astr\'ain, J.
Villadangos, C. Zubiri and F. Falcone, ``Characterization of Wireless Channel Impact on Wireless Sensor Network Performance in Public
Transportation Buses'', IEEE Transactions on Intelligent Transportation Systems, vol. 16, issue 6, pp. 3280-3293, 2015.

\bibitem{Salous2013} S. Salous, ``Radio Propagation Measurement and Channel Modelling''. Ed. John Wiley and Sons Ltd., 2013.

\bibitem{Loredo2008} S. Loredo, A. Rodriguez-Alonso and R. P. Torres, ``Indoor MIMO Channel Modeling by Rigorous GO/UTD-Based Ray Tracing'', IEEE Transactions on Vehicular Technology, vol. 57, no. 2, pp. 680-692, March 2008.

\end{thebibliography}
\end{document}